# Technological Understanding:

## On the cognitive skill involved in the design and use of technological artefacts

Eline de Jong[1*] & Sebastian De Haro[2]

**Abstract**

Although several accounts of scientific understanding exist, the concept of understanding in relation to technology remains underexplored. This paper addresses this gap by proposing a philosophical account of *technological understanding*: the type of understanding that is required for and reflected by successfully designing and using technological artefacts. We develop this notion by building on the concept of scientific understanding. Drawing on parallels between science and technology, and specifically between scientific theories and technological artefacts, we extend the idea of scientific understanding into the realm of technology. We argue that, just as scientific understanding involves the ability to explain a phenomenon using a theory, technological understanding involves the ability to use a technological artefact to realise a practical aim. Both theories and artefacts are tools, and using them successfully requires the cognitive skill of understanding. Technological understanding is thus conceived as the ability to recognise how a practical aim can be achieved by using a technological artefact. In a context of design, this general notion of technological understanding is specified as the ability to *design* an artefact that, by producing a phenomenon through its physical structure, achieves the intended aim. By analogy with De Regt's criterion of the intelligibility of theories, we give, as a precondition for technological understanding, a criterion for the intelligibility of a technological artefact. We illustrate our concept of technological understanding through two running examples: magnetic resonance imaging (MRI) and superconducting quantum computers. Our account highlights the epistemic dimension of engaging with technology and, by allowing for context-dependent specifications, provides guidance for testing and improving technological understanding in specific contexts.

[1,2] Institute for Logic, Language and Computation; Institute of Physics; Qusoft Research Center for Quantum Software; University of Amsterdam, Science Park 107 1098 XG, Amsterdam, The Netherlands

*Corresponding author: e.l.dejong@uva.nl, ORCID 0000-0002-9167-5585

**Conflict of interest**: All authors declare that they have no conflicts of interest.

**Funding**: This publication is an outcome of the project Quantum Impact on Societal Security, project number NWA.1436.20.002, which is funded by the Dutch Research Council (NWO), The quantum/nanorevolution.

**Acknowledgement:** We would like to thank Henk de Regt, Hans Radder, F.A. Muller, and Sonja Smets for their helpful comments on this manuscript. We also would like to acknowledge the constructive feedback received from participants in discussions at various conferences and workshops, including EASST-4S (July 2024), OZSW Annual Conference (August 2024), and Understanding Science & Technology (April 2024). These contributed significantly to the refinement of this paper.

# 1. Introduction

For several decades, 'understanding' has been recognised as a fundamental aim of science, alongside traditional aims such as prediction and explanation.[1] Although several accounts and developed theories of 'scientific understanding' exist, the question of understanding in relation to technology has remained relatively underexplored. One first line of work addresses the *knowledge* that is required for both designing and using technological artefacts (e.g. Houkes, 2009; Houkes & Meijers, 2022). However, these accounts do not discuss understanding as the ability to apply such knowledge. In a second line of work, understanding has been discussed in response to 'opaque' technologies[2] and in calls to improve understanding of new technologies to foster public debate (e.g. Vermaas, 2017). Recently, there has also been discourse on whether AI can act as an agent of scientific understanding (Barman et al., 2024; Krenn et al., 2022). In addition, Zhu (2025) calls for a hermeneutic notion of understanding to explore how users make sense of their interactions with technological artefacts and how such artefacts come to have meaning for them.[3] While these contributions valuably add to the emerging discussion, a comprehensive, epistemically informed philosophical account of what it means *to understand a technology* has yet to be developed.

The lack of such an account renders calls for improved understanding of specific technologies vague and potentially ineffective. Furthermore, without a clear explication of what it means to have understanding in relation to a technology, the cognitive dimension of our interactions with technological artefacts remains obscure. This paper aims to address this gap, by proposing an account of *technological understanding*: the kind of understanding that is involved in designing, reasoning about and using technological artefacts.

We develop our notion of technological understanding by drawing on Henk de Regt's (2017) account of scientific understanding. In this account, understanding is defined as a cognitive skill that enables one to apply scientific knowledge to perform specific tasks—most notably, explaining phenomena. This requires that the knowledge or theory in question is intelligible to the user. We argue that this conception of understanding offers a valuable framework for developing a similar notion in the realm of technology. Similar to the development and use of theory in science, the design and use of technological artefacts require an epistemic ability[4] to apply knowledge. This involves the cognitive skill of anticipating the consequences of the artefact's operation within a given context.

In Section 2, we first introduce De Regt's notion of scientific understanding. We then give our conception of a technological artefact, and argue that there is an analogy between scientific theories and technological artefacts in terms of the instrumental roles they play in their respective domains.[5] Just as science seeks to understand and explain natural phenomena through the formulation and use of theories, technology seeks to solve practical problems through the design and use of artefacts. We use this analogy as a heuristic to transfer the notion of scientific understanding to the technological domain, thereby developing an analogous—yet distinct—notion of technological understanding.

---

[1] This paper was submitted to Synthese in March 2024 and, due to factors outside of anyone's control, it was only accepted for publication in February 2026. In March 2025, we also made it available as a preprint (De Jong & De Haro, 2025a), together with a second preprint (De Jong & De Haro, 2025b). In these preprints, we first presented and developed our account of technological understanding. We are aware that later related work appeared in Westerblad (2026) and Zhu (2025).

[2] For example, the lack of understanding of how certain Artificial Intelligence (AI) models work has been problematised and has given rise to the pursuit of 'explainable AI' (XAI) (Barredo Arrieta et al., 2020; Samek & Müller, 2019; Goodman & Flaxman, 2017).

[3] An important difference between our account and Zhu's call for technological understanding is that, while for Zhu (2025: p. 122) the concept of technological understanding 'refers to the way in which we make sense of how to interact with a technical artefact' (i.e. it takes our interaction with a technical artefact as its primary focus), the object in our account is a practical aim. We will discuss this further in Section 2.

[4] We take 'an ability' to be the potential to perform an action (see also Miller, 2022, p. 470).

In Section 3, we extend De Regt's account into the realm of technology, to obtain our analogous notion of technological understanding. De Regt emphasises the pragmatic aspect of understanding, explaining it as the ability to use knowledge. We endorse this view and argue that using a technological artefact similarly involves a type of understanding: namely, it requires the cognitive skill to recognise how the artefact's operation (under varying circumstances) can achieve a desired outcome. Just as scientific understanding entails the ability to use a scientific theory to explain a phenomenon, we argue that *technological understanding entails the ability to use a technological artefact to realise an aim.* The analogy between the two types of understanding thus highlights the cognitive skill involved in using a tool (a scientific theory vs. a technological artefact) in achieving a goal (explaining a phenomenon in science versus realising a practical aim in technology). By analogy with De Regt's criterion of the intelligibility of theories, we also give a precondition for technological understanding: namely, a criterion for the intelligibility of a technological artefact.

From the outset, we stress that we use the term 'use' in a broad sense, as *employing something (here: a tool) for a purpose.* In this conception, 'use' can refer to mental activities (such as employing a theory to explain a phenomenon, or recognising how an artefact's capabilities could serve an aim) as well as practical activities (physically employing an artefact to achieve an aim).[6] This allows us to draw an analogy between using a theory and using a technological artefact, while also leaving room for interpretative flexibility: the ability to use a technological artefact may involve actual practical use, but it can also include thinking up its design or recognising potential applications (see Section 3.2). In other words, how we specify 'using an artefact' affects the scope of technological understanding.

Thus, depending on the specifications of 'use', we can define sub-types of technological understanding, which typically apply in certain contexts. In this paper, we specify the general notion of technological understanding for a specific context: the context of design. To have a design-type of technological understanding entails the ability to realise an aim by *designing* a technological artefact. This type of understanding emphasises the role of knowledge about physical phenomena as well as the cognitive skill that is required to make such knowledge productive to realise a practical aim.

---

[5] We do not present the analogy between scientific theories and technological artefacts—and by extension, between scientific and technological understanding—as evidence in itself, but as a heuristic (and not a foundation) for conceptual development. In our case, the analogy serves to transfer the structure of De Regt's account of scientific understanding into the technological domain, thereby helping us articulate a notion of technological understanding.

[6] Thus we endorse the standard practice in the literature on understanding, explanation, and models, of taking 'use' in the broad English meaning of the term: in the current context of explaining and understanding, it is used to express functionality, instrumentality, and being an epistemic tool. Consider the following examples of various kinds of epistemic use in Frigg (2023): a hammer is used by a doctor to rest reflexes (p. 393), formal tools are used (pp. 26, 53, 96), equipment and physical instruments are used (pp. 82, 84), linguistic terms are used (p. 138), the experimental method is used (p. 90), and data is used as evidence (p. 92). Morrison and Morgan (1999) attest to the same varied usage of the word 'use' in the context of models that are used for epistemic purposes (pp. 11-12, 16, 21, 23, 28). Finally, De Regt (2019) explains how Maxwell and Boltzmann used mechanical models for understanding (pp. 199, 205, 213): indeed, Maxwell's 'use' includes the *designing* of these mechanical models.

To illustrate our account, we apply it to two examples: magnetic resonance imaging (MRI) and superconducting quantum computers. MRI, a first-generation quantum technology, has been in practical use for decades, while quantum computing, a second-generation quantum technology, has yet to fully materialise. By examining these examples, we illustrate what it means to possess, or to lack, technological understanding.

By giving a detailed account of technological understanding, this paper also clarifies what it means to 'understand' a technology. Specifically, the notion of technological understanding elucidates the epistemic dimensions of designing and using technological artefacts, thus highlighting the cognitive skills that these activities require. In this way, our account acknowledges the active, problem-solving nature of technological practice, where the success of understanding is not just measured by the ability to explain, but by the ability to realise practical outcomes. Furthermore, our framework moves beyond existing accounts by providing a way to specify understanding in relation to different technological contexts, thus enabling more precise approaches to fostering understanding of both established and emerging technologies.

## 2. Setting the Stage

In order to extend the concept of scientific understanding into the realm of technology, we introduce three *dramatis personae*: namely, a conception of scientific understanding (Section 2.1), an account of technological artefacts (Section 2.2), and an analogy between scientific theories and technological artefacts (Section 2.3).

### 2.1. Scientific understanding

In addition to scientific explanation, scientific understanding has been distinguished as a distinct cognitive ability, different from e.g. knowledge, and as a central aim of science (see for example: De Regt and Dieks, 2005; De Regt, 2009; Boon, 2009; Strevens, 2013).[7] According to what Reutlinger et al. (2018, p. 1081) call 'typical accounts of scientific understanding', scientific understanding requires that there is a (true) scientific explanation of a phenomenon, and that this explanation is *epistemically accessible*.[8] In such accounts, a phenomenon $P$ is understood by a scientist $S$ if and only if there is an adequate theoretical explanation for $P$ that is epistemically accessible to $S$. This raises the key question: What does it mean for an explanation to be epistemically accessible?

[7] For a discussion and overview of this 'typical account' of scientific understanding and its relation to explanation, see De Haro and Butterfield (2025: pp. 535-536).

[8] There is a debate in the literature about whether scientific understanding is objective, or whether it is a subjective concept, of no interest for the philosophy of science. Thus in an oft-quoted passage, Hempel (1965, p. 413) writes that 'such expressions as 'realm of understanding' and 'comprehensible' do not belong to the vocabulary of logic, for they refer to the psychological or pragmatic aspects of explanation'. However, Hempel (1965, p. 337) can also be read as holding a more moderate eliminativist view, that 'explanation is understanding enough' (for a summary of this debate, see De Haro and Butterfield (2025: pp. 531-534)). Recently, a number of authors have also distinguished between the use of 'pragmatic' as 'subjective', and the broadly Wittgensteinian sense of 'useable for certain aims'. This sense is compatible with an objective (appropriately inter-subjective) notion of understanding. We here endorse this consensus: see e.g. Friedman (1974, pp. 18-19), Salmon (1978, p. 684), Kitcher (1981, pp. 509, 529; 1989, p. 419), Schutz and Lambert (1994, p. 66), Lipton (2004, p. 30), De Regt and Dieks (2005, p. 150), Grimm (2010, p. 337), and Reutlinger et al. (2018, p. 1081).

De Regt and Dieks (2005), and later De Regt (2009; 2013; 2015; 2017), offer a well-developed theory to address this issue. De Regt defines the epistemic accessibility of an explanation—and thus, scientific understanding—as the *ability to use a theory* to give an explanation (2005, p. 142). This leads De Regt (2017) to formulate the Criterion for Understanding Phenomena (CUP):

> **CUP**: *A phenomenon $P$ is understood scientifically if and only if there is an explanation of $P$ that is based on an intelligible theory $T$ and conforms to the basic epistemic values of empirical adequacy and internal consistency.*[9] (2017, p. 92)

In this criterion, the epistemic accessibility of an explanation is given by the *intelligibility* of the theory used. De Regt defines intelligibility as 'the value that scientists attribute to the cluster of qualities of a theory … that facilitate the use of a theory' (2017, p. 40). Thus, for scientists to use a theory to explain phenomena, the theory must be intelligible to them. As one way (among others) to objectively assess this, De Regt proposes a Criterion for Intelligible Theories (CIT):[10]

> **CIT**: *A scientific theory $T$ (in one or more of its representations) is intelligible for scientists (in context $C$) if they can recognise qualitatively characteristic consequences of $T$ without performing exact calculations.*[11] (2017, p. 102)

Scientific understanding is thus explained as qualitative insight into the consequences of a theory—that is, a skilled recognition of those consequences.

It should be noted that this intelligibility criterion 'captures the pragmatic and contextual nature of intelligibility (and accordingly of understanding)' (De Regt & Dieks, 2005, p.151), and allows for variation in standards of intelligibility—and by extension, of achieved understanding—among scientific communities. De Regt emphasises that intelligibility is not an intrinsic property of a scientific theory. Rather, it depends on both the characteristics of the theory and on the relevant scientific community that uses it. In other words, the intelligibility of a theory requires a match between its theoretical virtues and the cognitive skills, background knowledge, problem-solving practices, and aims of the scientific community using it. This makes intelligibility a *contextual* notion: it cannot be assessed independently of the context in which a theory is used. This implies that scientific understanding is context-dependent too: if the intelligibility of a theory depends on the user context, then scientific understanding—defined as the ability to use that theory to explain phenomena—also depends on this context.

By emphasising the role of the cognitive and epistemic skills that enable the use of a theory, De Regt's account highlights the *pragmatic* character of scientific understanding. Thus, in his view, understanding is not merely the possession of knowledge, but the ability to actively use knowledge[12] to explain phenomena—it is an 'extra cognitive ingredient' that is required to use a theory successfully (De Regt & Dieks, 2005, p. 149). This involves cognitive skills such as making explanatory connections, answering what-if-things-had-been-different-questions, and providing clarifying examples (see also De Regt's latest elaboration in Barman et al., 2024). For example, beyond knowing that there is global warming and that this is mainly caused by the increased atmospheric levels of carbon dioxide, scientific understanding of climate change entails the ability to explain this causal relationship by using an intelligible theory.

[9] The notion of scientific understanding has frequently been discussed in relation to the deductive-nomological (D-N) model of scientific explanation. According to this model, notably advocated by Hempel (1965), a scientific explanation is a deductive argument that derives the occurrence of a phenomenon $P$ from general laws of nature in conjunction with specific conditions. De Regt himself adopts a pluralistic stance on scientific explanation, asserting that his account of scientific understanding is compatible with various models, including causal, contextual, and non-deductive approaches. This pluralism acknowledges the complexity of scientific practice and recognises that different contexts may require different explanatory frameworks.

[10] Boon (2009, p. 6) has criticised De Regt's CIT, arguing that intelligibility is not best defined as the ability to recognise qualitatively consequences of $T$; intelligibility can also be defined more broadly, as the ability to use $T$ in reasoning to solve relevant problems.

[11] This formulation leaves room for variation of standards of intelligibility, and hence, of scientific understanding: for example, across scientific communities. This does not, however, make intelligibility a subjective notion. For, even though its standards can vary, the criterion itself does not change, and it functions as an objective test for theories (see also footnote 4, about the objectivity of scientific understanding).

This pragmatic character of De Regt's account of scientific understanding, where understanding is defined as the ability to actively use knowledge to perform certain tasks (Barman et al., 2024), makes it particularly relevant to our project of developing a similar notion within the practical and goal-oriented domain of technology (see also Sections 2.2 and 2.3).[13] Tentatively, it appears that the design and use of a technological artefact also requires an epistemic ability to use knowledge, including a cognitive 'intuition' for the consequences of the artefact's operation, given a certain context. Furthermore, the contextual notion of intelligibility—defined as relative to both the aims of the users and the skills they possess—aligns well with the practice of using technological artefacts.

To extend the concept of scientific understanding into the realm of technology, we begin by discussing the technological counterpart of a scientific theory: namely, the technological artefact.

## 2.2. The conception of a technological artefact

Just as scientific theories are usually taken as the core units of analysis in science, we will take technological artefacts to be the primary units of our analysis of technology.[14] Like scientific theories, technological artefacts are constructed or *designed* objects. Note that what makes artefacts *technological* is their inherently 'functional', 'useful', and 'instrumental' nature (Kroes, 2002, pp. 292-294; Houkes & Vermaas, 2010; Van de Poel, 2009, p. 980). Hughes even speaks of the 'instrumental function' of artefacts, to highlight that they are designed and valued for their effectiveness in performing tasks or solving problems (Hughes 2009, p. 181). In this paper, we take technological artefacts to be *designed, functional objects*.[15]

To develop a nuanced notion of understanding in relation to technology, particularly when considering more complex technologies, it is helpful to explore the functional nature of artefacts in greater detail. To this end, we draw on the work of Kroes (2002).

[12] In the rest of this paper, our usage of 'knowledge' in the context of technological understanding will encompass a combination of causal knowledge and factual knowledge of laws, theories, technological and design procedures, and background conditions (see Section 2.1).

[13] As noted in the introduction, Zhu (2025) calls for a hermeneutic notion of understanding in relation to technology that emphasises the way in which we 'make sense' of our interactions with technological artefacts. Our account rather focusses on the epistemic and pragmatic dimension of understanding in relation to technology (see also footnote 2).

[14] We use 'technology' in an abstract sense, not referring to a particular object, and 'technological artefact' as a concrete object with a specific function.

[15] Defining technological artefacts as *functional* objects in terms of realising practical aims does not deny that functional requirements are not the only considerations that shape an artefact (see Van de Poel, 2009, pp. 986-987 about 'additional requirements'), nor does it exclude the related possibility of artefacts having functions that we would call symbolic rather than practical—although a practical function is essential in our conception of technological artefacts.

Kroes (2002) explains the functionality of technological artefacts through two key aspects: (i) an intentional aspect and (ii) a physical aspect. The intentional aspect is the function attributed[16] to an object, shaped by human intentions. The physical aspect comprises the material properties and structure that make it possible to fulfil this function. We endorse this dual-aspect view, and define a 'technological artefact' as a *functional object* that possesses both *intentional and physical aspects.*

To deepen our understanding of the *physical aspect*[17], we break it down into two components: a *physical structure* and a *physical phenomenon*:

(i) *Physical structure,* $X$: This is the physical construction of the artefact, capable of producing a specific phenomenon as the result of its components working together. It is the material parts—like the metal head of a hammer or the chips in a computer—that make functionality possible.

Since the physical structure consists of parts that work together, it will be useful when analysing a particular artefact to denote these by $x_1, x_2$, etc. For example, in magnetic resonance imaging (MRI), we can distinguish between the main superconducting magnet that polarises the sample of tissue (i.e. $x_1$), the shim coils that correct the shifts in the homogeneity of the magnetic field (i.e. $x_2$), the gradient system that is used to localise the region of tissue that is being scanned (i.e. $x_3$), etc.

In the case of a superconducting quantum computer, the physical structure consists of several key components. The main superconducting circuits (i.e. $x_1$), known as the quantum processing units, house the superconducting qubits ('quantum bits', i.e. quantum analogues of classical bits) or transmons, which themselves have sub-components like a Josephson junction, a capacitor for external interaction through photon absorption, etc. The control electronics (i.e. $x_2$) generate precisely timed microwave pulses that manipulate the states of the qubits, and the cryogenic systems (i.e. $x_3$) maintain the ultra-low temperatures necessary for ensuring qubit stability and minimising thermal noise, which can disrupt the fragile quantum states. Other components include coaxial cables, which transmit signals with minimal loss, shielding from external electromagnetic interference, etc.

(ii) *Physical phenomenon,* $P$: This is the functional phenomenon resulting from the operation of the physical structure. It is the falling weight of the hammer or the electric signals running through computer chips.

In our example of MRI, the phenomenon is the production of the digitalised image of the region of tissue being scanned. As a *phenomenon,* this image is an uninterpreted material object, produced by a series of physical and chemical operations that follow the MRI's operations. Like with the physical structure, we can distinguish several constitutive sub-phenomena within the phenomenon $P$, like the creation of the magnetic field with particular properties of homogeneity (i.e. $p_1$), the polarisation of the protons in the sample of tissue (i.e. $p_2$), the resulting magnetic resonance signal (i.e. $p_3$), etc.

---

[16] Kroes (2002) distinguishes a context of *design* and a context of *use*, both of which can be constitutive of the technological function. In this paper, we focus on the intentional context of design and hence on the functionality that is defined and assessed at this stage.

[17] The physical structure includes seemingly non-physical objects, like software, which are realised with various types of physical supports. In this paper, we focus on the initial or fundamental (that is: material) design process, which explains the focus on the physical aspect of technological artefacts. With appropriate modifications, and using the analogy between technological artefacts and scientific theories, we believe that our framework can also be adapted to the study of data-based design processes such as software engineering.

In the case of a superconducting quantum computer, the central phenomenon is the behaviour of qubits, which are realised as superconducting circuits or transmons. This behaviour can be analysed into several phenomena. One such phenomenon is quantum tunnelling (i.e. $p_1$) where pairs of electrons (Cooper pairs) tunnel through a Josephson junction. This enables the formation of unevenly spaced discrete energy levels (i.e. $p_2$), thus allowing the selection of the lowest two states as qubit states, namely the ground state and the first excited state. Another is superposition (i.e. $p_3$), where each transmon qubit is in a combination of its ground and excited states simultaneously, enabled by the controlled oscillations in the superconducting circuit. Entanglement (i.e. $p_4$) also plays a vital role, where multiple qubits become correlated through shared electromagnetic interactions, allowing for the complex interactions necessary for quantum computation. These are just a few key phenomena; others include quantum measurement, (de)coherence, quantum error correction, etc.

The *intentional aspect* of a technological artefact refers to its relation to a specific aim: the technological artefact $t$ realises, achieves, or embodies its intended aim $A$ (see also Weber et al., 2013). For a technological artefact to realise an aim typically means to solve a practical problem or fulfilling a specific need of a group of (prospective) users. This can be virtually any type of practical need: from personal transportation to communication over distance, and from solving computational problems to providing heat and light. We refer to the aims of technological artefacts as 'practical', since they require the artefact's practical use.[18]

We further break down the intentional aspect into two levels: *direct aim* and *ultimate aim*:

(i) *Ultimate aim,* $A$*:* This is the ultimate effect that the artefact is intended to achieve.[19] The ultimate aim is 'external' to the artefact, in the sense that it can exist independently from the specific artefact, and a given technological artefact can accommodate for various ultimate aims (see also Kroes, 2002).

In the example of MRI, the ultimate aim is to perform a medical examination of organs, tissues, and the skeletal system in a non-invasive way. For a quantum computer, the overarching ultimate aim is to solve a specific class of computational problems that are (practically) intractable for classical computers. Achieving this aim potentially enables a wide range of applications, from accelerating drug discovery and the development of new materials, to optimising logistics and simulating complex natural systems.

(ii) *Direct aim,* $a$: This is the specific way the technological artefact achieves—or approximates—its ultimate aim through its operation. It is the specific technological response to the ultimate aim—for example, like performing an MRI scan to detect tissue abnormalities is one way to (technologically) support medical diagnosis. The direct aim is implicit in the production of the phenomenon; they are two sides of the same coin. The direct aim adds a layer of interpretation of the produced phenomenon in terms of intentionality. We make this explicit in our notion by writing the technological artefact $t$ as a triple: $\langle X, P, a \rangle$.

[18] By characterising the aim of a technological artefact as *practical*, we do not wish to contrast 'practical' and 'theoretical'; since in the previous Section we already said that artefacts can be used for epistemic aims (think of e.g. telescopes and particle accelerators). Rather, 'practical' here indicates that the realisation of the aim requires the artefact to do something, i.e. to perform a task that leaves the artefact and its immediate environment in a different state that it was before.

[19] A technological artefact can—and often will—have other effects and uses than initially foreseen or intended during the design process. In this paper, we limit our scope to the intended use of an artefact and its associated effect.

The ultimate aim of the MRI can be technologically realised by the MRI's direct aim of producing medical images of the inner human body according to set specifications, i.e. images that are interpreted by a doctor. In the case of the quantum computer, the ultimate aim of unlocking increased computational abilities can be realised through the direct aim of executing specific quantum algorithms on a fault-tolerant quantum computer—that is, a quantum computer equipped with a sufficient number of reliable qubits that possess adequate coherence times, dependable quantum gate operations, robust error correction mechanisms, and high measurement fidelity.

The distinction between the artefact's direct and ultimate aim is useful for two reasons. First, it allows us to distinguish between the aim as it is instantiated by the artefact (namely the direct aim, e.g. generating a medical image with a specific resolution quality), and the aim that can be conceptualised independently of the specific artefact (the ultimate aim, e.g. achieving improved medical diagnosis). The distinction between the ultimate aim $A$ and its embodiment $a$ in a technological artefact, is analogous to the distinction between a phenomenon $P$ and the description of that phenomenon given by a scientific theory: A scientific theory provides a particular description or perspective on $P$; in a similar way, the direct aim articulates a particular interpretation of the ultimate aim $A$. Both a theory and an artefact's direct aim thus function as mediating descriptions that connect a broader object (a phenomenon or an ultimate aim) to a concrete way of engaging with it.

Second, it highlights that $a$ is often an approximation of $A$. As Van de Poel (2009, p. 985) emphasises, we usually lack an overview of the complete set of possible solutions to a design problem (i.e. the problem for which the artefact-to-be-designed is supposed to offer a solution). The aim $a$ captures only one way of operationalising the ultimate aim $A$, and could even do so partially. In this sense, $a$ represents a solution to $A$ that is 'good enough' given the circumstances. We use the term 'approximation' to emphasise the open-endedness of specifying $A$ in practice: each direct aim reflects a context-dependent—and often partial—way of realising the ultimate aim.

As we already pointed out, it will be clearest to consider the ultimate aim as *external* to the artefact. That means that the ultimate aim $A$ is, strictly speaking, not part of the technological artefact $t$, i.e. the triple $\langle X, P, a \rangle$. While a technological artefact $t$ requires *some* ultimate aim $A$ to be considered functional, it does not need to be tied to a *specific* ultimate aim from the outset. Furthermore, the ultimate aim of an artefact can change over time. Although one could argue that (the meaning or character of) an artefact changes depending on what it is used for, we maintain that the artefact itself is the same for various ultimate aims (so long as $X, P$ and $a$ do not change). Thus fixing or determining a *specific* ultimate $A$ is not required to define a technological artefact. In this way, the technological artefact and the ultimate aim are logically independent, and they are related by the degree to which, once fixed, the direct aim matches the ultimate aim. Therefore, the question of the degree to which a technological artefact $t$ serves an ultimate aim $A$ is a normative question about how well the direct aim $a$ realises or approximates the ultimate aim $A$.

Our conception of a technological artefact, i.e. $t = \langle X, P, a \rangle$, is summarised as follows:

> *A* ***technological artefact*** *is a designed object, consisting of a physical structure $X$ that produces a physical phenomenon $P$ to achieve a direct aim $a$, which serves an ultimate aim $A$.*

This 'assembled' conception of technological artefacts—defining it as an interplay of physical phenomena, physical structure, and intended functionality—is embedded in extant accounts by Kroes (2002), and Houkes & Vermaas (2010), and Boon's work on the role of physical phenomena in the engineering sciences (2017). Our aim is not to present a wholly novel account of artefacts, but to foreground and integrate specific aspects into a single, multifaceted definition. By explicitly including the role of phenomena in attaining an aim, and distinguishing between direct and ultimate aims, our conception allows us to demonstrate how understanding in different contexts may focus on different components of the artefact. Additionally, our conception underscores how scientific understanding underpins technological artefacts—an emphasis particularly appropriate for advanced technologies that rely heavily on the production and manipulation of certain physical phenomena, such as the polarisation of nuclear spin or quantum entanglement which are at the core of quantum technologies.

In the next Section, we use this conception of technological artefacts as a conceptual building block to develop an analogy between science and technology and, more specifically, between scientific theories and technological artefacts.

## 2.3. An analogy between scientific theories and technological artefacts

To extend the concept of understanding from the scientific realm into the realm of technology, we must first explore the relationship between these two fields.

Science and technology have a 'symbiotic' relationship (Barnes 1982, p. 168), in which they inform and shape one another. Just as scientific breakthroughs lead to the development of new technologies, technological advancements also often enable new scientific discoveries (Boon, 2015). Thus technological artefacts can be seen both as outcomes of scientific understanding and as contributors to its advancement.

This mutual relationship is grounded in a shared concern with understanding physical phenomena. In the uses of both a theory and an artefact, a phenomenon is 'generated' as an outcome: respectively conceptually through theoretical derivation, explanation and prediction, and practically through the operation of the artefact. Furthermore, in experimental science, these two processes go together when phenomena are practically produced in the lab.

The aims of their respective concerns, however, differ (Boon 2006; 2009). In science, the concern with phenomena is first and foremost *epistemic* (De Regt, 2009): the primary goal is to explain and understand phenomena, with scientific theories and models serving as our tools for doing so. In technology, the concern with understanding phenomena is *instrumental*: the goal is to 'exploit' phenomena (i.e., make them productive) to achieve practical aims, using technological artefacts as tools (see also Boon, 2020: pp. 80-94). However, these aims can be considered deeply intertwined: explanatory information is information that is (potentially) relevant to manipulation and control, and vice versa (Woodward, 2003).

This perspective highlights that the technological exploitation of phenomena is always based on an underlying explanation, whether implicit or explicit (see also Woodward, 2003; Boon, 2006, p. 27). In this sense, technology invariably builds on a—contextual, and more mundane or more

advanced[20]—scientific understanding of phenomena, somewhere down the line. For example, designing a laptop relies on scientific understanding of the semiconducting properties of transistors in processing chips. Similarly, designing a quantum computer depends on an advanced grasp of quantum phenomena like superposition and entanglement. Every technological artefact uses and takes advantage of scientific understanding of phenomena and makes it productive for some practical aim.

Technological artefacts can thus be considered (physical) instantiations of scientific knowledge and explanations. In that respect, artefacts can be seen as analogous to scientific theories: where a scientific theory can be used to explain a phenomenon abstractly, a technological artefact operationalises the explanation[21] in a concrete, material form, thus 'embodying' the theory.[22] The use of a technological artefact can be thought of as a special instance of using scientific knowledge, which is why the concept of scientific understanding (as the ability to use a theory) offers an interesting starting point for conceiving a notion of understanding in the context of technology (Boon, 2015).

The analogy between scientific theories and technological artefacts thus centres on their engagement with physical phenomena. Scientific theories and technological artefacts are analogous in that both function as tools bridging physical phenomena and human aims: theories provide explanations of phenomena (to achieve understanding), while artefacts exploit these phenomena (to realise practical outcomes).

Similarly, Morgan & Morrison (1999) argue that there is an analogy between (theoretical) models and technology. They emphasise that models function as tools or instruments, that facilitate recognising how a theory is applied in specific circumstances (pp. 10-11). In that sense, they argue, a model holds the quality of a technology, demonstrating its power in its use. Furthermore, they stress that models can work as useful instruments for the design of technologies since 'they provide the kind of information that allows us to intervene in the world.' (p. 23) While artefacts can indeed be perceived as akin to models in their capacity to bridge the gap between theory and practice, we advocate for a more direct analogy between scientific theories and technological artefacts, endorsing the view that theories serve as foundations of models, and that artefacts typically operationalise general theoretical principles (see, in particular, De Regt, 2017, pp. 31-36).

However, this is not to say that scientific understanding is sufficient for the design of technological artefacts (cf. Zhu, 2025). As Boon (2006) notes, 'using science' in artefact design requires 'substantial additional work.' This additional work relates to the nature of technological artefacts, which we previously defined as objects that are capable of producing specific phenomena through a physical structure to achieve a particular aim. Understanding the functional phenomenon is only one aspect of the process; other aspects include choosing an appropriate phenomenon given the particular aim, and thinking up a physical structure that can produce that phenomenon. This distinction hints at the differences between theories and artefacts, which we will explore further in the following Section.

[20] The degree to which a technological artefact builds on scientific understanding depends on the complexity of the technological artefact.
[21] While there are various models of explanation relevant in science, the ones that are most relevant to the design and use of technological artefacts are the causal and the mechanistic models (see also Boon, 2006, p. 27).
[22] For a recent account of the usages of 'theory' and 'model', see Frigg (2023).

Given this shared focus on understanding physical phenomena, and the way technology builds on scientific explanations to exploit those phenomena, we propose an analogy between science and technology, particularly between scientific theories and technological artefacts. While there are important similarities between theories and artefacts, there are also key differences—hence the analogy, rather than identity. The analogy primarily serves to validate the relevance of employing the concept of scientific understanding in developing a notion of understanding in the context of technology. As we discussed in Section 2.1, this relevance predominantly stems from the notion of scientific understanding as a cognitive skill to perform certain actions. Thus ultimately, our conception of technological understanding is intended to stand on its own, independent of the analogy that we use to construct it. Its usefulness will be illustrated by how it clarifies what it means to understand technology in different contexts.

# 3. A Conception of Technological Understanding

In this Section, we develop our notion of technological understanding, in analogy with scientific understanding. We first give a general conception of technological understanding (Section 3.1), followed by a discussion of the scope of this concept (Section 3.2), and a specification of technological understanding for a context of design (Section 3.3).

## 3.1. Technological understanding: the ability to use a technological artefact

As we already discussed in Section 2.3, there is an analogy between scientific theories and technological artefacts. For example, both serve as an *instrument* to achieve specific aims: scientific theories explain phenomena, i.e. answering the question: *Why (does) P (occur)?*, while technological artefacts are designed to realise practical aims, i.e. solving a problem or meeting a need. This analogy was strengthened by the idea that a technological artefact operationalises the explanation of a phenomenon, making it productive in its operation. By doing so, the artefact 'exploits' the phenomenon to achieve its intended function.

We use this analogy between scientific theories and technological artefacts as a stepping stone for transferring the notion of scientific understanding to the context of technology.[23] If we think of understanding as the cognitive skill to (recognise how to) use an instrument to achieve an aim in varying circumstances, then successfully using a technological artefact indeed involves a type of understanding. Building on the concept of scientific understanding as the ability to use a theory to explain a phenomenon, we propose *technological understanding* analogously as the ability to use a technological artefact to realise a practical aim.

It is important to note that 'the ability to use a technological artefact to realise an aim' is the conception of technological understanding in its most generic form: the 'use' of a technological artefact is understood here broadly, and includes design and (conceptual as well as actual) use. This broadness mirrors De Regt's notion of scientific understanding, which applies in both the context of developing a new theory, and in the context of using an existing theory to explain a particular phenomenon. In the next Section, we will explore in more detail how this general conception of 'use' can be scoped and refined. For now, however, we leave 'use' open to further specification.

[23] In this sense, the concept of technological understanding should ultimately stand on its own merits, independently of the analogy.

In this generic sense, we describe the ability to use a technological artefact as the capability to provide an appropriate 'technological response' to a specific aim—one that successfully achieves the desired outcome. This may involve designing, operating, or conceptualising the artefact (see Section 3.2), and it requires insight into the need or demand itself as well as into how it can be technologically addressed. Thus technological understanding is the cognitive skill that is both demonstrated in, and necessary for, the successful use of a technological artefact.

We can explain this cognitive skill as an ability to use knowledge of the artefact's operation flexibly across different contexts. In other words, having technological understanding means being able to reason about how an artefact can be used to achieve a given aim. Like scientific understanding, technological understanding involves explanatory reasoning. However, while scientific understanding does not privilege a particular dimension of explanation, technological understanding stresses the causal dimensions of explanation. Because its purpose is to bring about a particular state of affairs, technological understanding requires recognising how the artefact's operation leads to the realisation of the intended outcome.

One demonstrates technological understanding when one can foresee how an aim can be achieved through the use of a technological artefact: for example, understanding how an umbrella can be used to stay dry during a rain shower while anticipating that strong wind may reduce its effectiveness. In the case of an MRI machine, technological understanding involves recognising how it can be used for non-invasive medical examinations. Similarly, for a superconducting quantum computer, technological understanding would involve the ability to recognise how to use it for performing the kind of computations that, for example, could accelerate drug discovery.

Defining technological understanding as the ability to realise an aim by using a technological artefact implies that the *object* of technological understanding is the ultimate aim $A$. Having technological understanding of an aim means recognising how it can be achieved by the operation of a technological artefact (i.e. $t$). Thus the matter to be technologically understood is not the artefact itself but a desired outcome—just like it is the phenomenon that is the object of scientific understanding and not the theory itself. However, since technological understanding involves recognising how the artefact's operation realises an aim, it requires a level of 'understanding' of the artefact itself. In other words, technological understanding requires that the technological artefact is *intelligible*. While technological understanding is, strictly speaking (a specific kind of) understanding of an aim, it could also be informally described as 'understanding a technology'.

Building on De Regt's CUP for scientific understanding (see Section 2.1), we propose a Criterion for Technological Understanding (CTU):

**CTU:** *An aim $A$ is technologically understood if it can be realised by using a technological artefact $t$.*

Applying this criterion to the MRI example, we see that technological understanding is achieved: the aim of non-invasive medical examination is realised through the use of the MRI machine. By contrast, there is not yet full technological understanding of the aim to accelerate drug discovery through a quantum computational advantage,[24] since a fully functioning quantum computer does not yet exist to achieve this aim.

To achieve scientific understanding of a phenomenon $P$, one must select or develop a theory that is capable of adequately explaining it. Similarly, technological understanding of an aim requires selecting an artefact that can successfully achieve the desired outcome—or, if such an artefact is not available, designing it (see Section 3.3). Selecting an adequate artefact is implied by having technological understanding.

Since the ability to use a technological artefact requires that the artefact is intelligible to the user, we also propose a Criterion for Intelligible Technological Artefacts (CITA), inspired by De Regt's CIT:[25]

> **CITA:** A technological artefact $\mathsf{t}$ is intelligible for a subject $\mathsf{S}$ (in context $\mathsf{C}$) if $\mathsf{S}$ can recognise qualitatively characteristic consequences of $\mathsf{t}$'s operations without practically performing these operations

The ability 'to recognise qualitatively characteristic consequences' can be explained in terms of the ability to explore the space of possible initial and end-states of the artefact without practically performing its operations (and regardless of whether, and how, the ultimate aim $A$ is realised). Such exploration requires cognitive skills and reflects the epistemic activity that is involved in technological understanding. This can be characterised as a qualitative recognition of the consequences of the artefact: a skilled judgement of how it works and how it might be used.

CITA's explicit reference to the subject[26] $S$ and the context $C$ points to the pragmatic and contextual character of the artefact's intelligibility, and so, of technological understanding. The intelligibility of an artefact cannot be established in isolation, but only in reference to a (typical) subject: the same artefact might be intelligible to one (type of) user, but a black box to another. Intelligibility thus depends on the match between the artefact and contextual factors such as skills, familiarity, and the role the user: whether the artefacts is deemed (sufficiently) intelligible can only be assessed in reference to the subject *to whom* it is (un)intelligible. In other words, intelligibility is contextual and allows for variation among different groups of users, which in turn leads to variation in technological understanding.[27]

To illustrate, consider the MRI machine. A medical technician or radiologist understands the MRI's operation well enough to anticipate how adjustments in magnetic fields and radiofrequency pulses will produce different types of images, such as distinguishing between soft tissues or identifying specific abnormalities. Even without running every possible scan, they can predict the consequences of various settings and inputs. This ability to foresee outcomes, based on knowledge of the machine's functioning, makes the MRI intelligible to them—and lacking this ability render the MRI unintelligible to many others.

[24] A 'quantum advantage' is achieved when the increased computational power enables quantum computers to solve mathematical problems that classical computers in practice cannot solve (Hoofnagle & Garfinkel, 2022).

[25] To flag that there may be other ways to determine the intelligibility of a theory, De Regt (2017, p. 102) adds the subscript $CIT_1$ to his criterion, and explains: '$CIT_1$ is a sufficient condition for intelligibility, not a necessary one; there may be alternative criteria $CIT_2$, $CIT_3$, etc.' Although, for simplicity of the notation, we will not use the subscript in this paper, we are equally open to the possibility of testing the intelligibility of a technological artefact in other ways.

[26] Our subject $S$ stands for a typical subject of the relevant community, and is thus contextually determined.

[27] This parallels De Regt's pragmatic, context-dependent conception of scientific understanding, which likewise allows for differences in standards of intelligibility of theories, and hence in scientific understanding.

In contrast, with the superconducting quantum computer, while there is a basic understanding of its principles (see Section 2.2), there is not yet a fully intelligible artefact, simply because the design has not yet been successful. There may be scientific understanding of the physics underpinning quantum computing (De Regt & Dieks 2005; De Regt 2017),[28] but the lack of practical implementation in a large-scale, fault-tolerant quantum computer with reliable outcomes means that in this case the artefact's operation and its consequences are not yet recognisable. We delve further into this in Section 3.3, where we explore technological understanding within a design context. For now, it is important to see that the absence of an intelligible artefact implies a lack of technological understanding in the case of superconducting quantum computers.

The concept of technological understanding illuminates the cognitive dimension of designing and using technological artefacts. By highlighting that these activities involve cognitive skills and reasoning with imperfect knowledge, it shows that the use and design of technological artefacts require the ability to effectively navigate and use knowledge. Furthermore, technological understanding itself emerges as an epistemic aim, critical to the successful use of artefacts. From this perspective, the successes, challenges, and failures in designing and using artefacts can be seen as cognitive achievements or limitations, offering a richer framework for analysing technological practice.

## 3.2. Specifying technological understanding: three logical options

Having outlined a general conception of technological understanding, we now turn to exploring the logical options for its further specification. As we mentioned in the previous Section, the phrase 'ability to use a technological artefact' has so far remained vague, particularly regarding the meaning of 'use': Does it include design, and does it imply the actual operation of the artefact? In this section, we explore three key options for specifying technological understanding. Doing so will eventually help to clarify what technological understanding entails for typical groups of users, representative of a broader category of subjects.

We base the available options on whether the aim and the artefact are determined, since this affects the cognitive requirements of technological understanding and, ultimately, how 'use' is specified. Depending on whether the aim and the artefact are predetermined (i.e. 'fixed') or underdetermined (i.e. 'open'), we identify three logical options for an account of technological understanding:

(i) *Maximal account:* where both the aim $A$ and the technological artefact $t$ are open.

(ii) *Minimal account*: where both the aim $A$ and the technological artefact $t$ are fixed.

[28] Indeed, the debate between Heisenberg and Schrödinger over the *Anschaulichkeit* of quantum theory is one of the major historical case studies to which De Regt's theory applies (see De Regt, 2017, Chapter 7). We endorse the clear verdict of the case studies, that quantum phenomena, including those, like superposition, that play a role in quantum computation, *can* be scientifically understood (to a satisfactory degree) by using quantum mechanics. This verdict disagrees with those conceptions of understanding that require additional conditions (for example, visualisability, or a specific metaphysics based on classical theories, etc.), and that we argue are only reactionary attempts to retain a notion of understanding that is moulded on specific superseded theories. For when it comes to understanding quantum phenomena, the more open-minded positions in the neighbourhood of Pauli's and Heisenberg's views, rather than Schrödinger's view of understanding as *Anschaulichkeit*, are vindicated. For a discussion, see De Haro and Butterfield (2025: pp. 119-120) and De Haro and De Regt (2020: p. 3122).

(iii) *Via media*: where the aim $A$ is fixed, and the technological artefact $t$ is open; or where the aim $A$ is open and the technological artefact $t$ is fixed.

The logical strength, or scope, of these accounts differs with respect to what technological understanding encompasses. In the *maximal account* (i), technological understanding involves both selecting or formulating the aim $A$ and devising a technological artefact $t$ that achieves this aim. We refer to this as the 'maximal' account because it involves the greatest degree of cognitive freedom. This specification of 'use' typically applies to situations where a new connection between an aim and a (prospective) artefact is invented or devised. In this option, the use of an artefact can thus be specified as 'conceptualising' the artefact.

In the *minimal account* (ii), technological understanding does not involve selecting either the aim or the artefact. Instead, it involves enacting an already available connection between the aim and the artefact. This applies to the common employment of the artefact. In this case, 'use' is specified as actual operation, but also in the form of the ability to give an explanation of how the artefact can be operated to achieve the aim. In other words, for this type of technological understanding, one requires an 'explanation of the output from the input and the aims' by reasoning—either explicitly or implicitly by doing.

The *via media* (iii) is the intermediate option between the maximal and the minimal accounts. Here, technological understanding involves determining or 'thinking up' the artefact to achieve a given aim, or finding a new aim for an already available artefact. The first scenario typically applies to situations where an artefact is designed to meet a specific, predetermined aim. In this option, 'use' is specified as 'design for use'.

We consider the second scenario (given artefact, open aim) to be a variant of the maximal account (i). In that account, the emphasis is on thinking up the relation of functionality between an artefact and an aim—which closely resembles the process of instrumentalising an existing artefact for a new aim in this second scenario of the via media. In other words, 'repurposing' an artefact can be considered an instance of 'conceptualising' it. Therefore, we do not consider this scenario as a separate option.

By contrast, we treat the scenario of 'open artefact, given aim' as a distinct option because it directly engages with the practice of design. Here, the artefact must be conceived to meet a given aim, which, as we will argue in Section 3.3, requires a unique form of technological understanding that emphasises cognitive freedom in constituting the artefact.

Thus, the three logical options for specifying 'the ability to use an artefact to realise an aim' yield three corresponding specifications or subtypes of technological understanding. Crucially, each subtype assumes that the artefact is intelligible to its user (understood here as a representative group of subjects): it is this intelligibility that makes technological understanding possible in the first place. Yet, just as standards of intelligibility for theories may differ across scientific communities, the intelligibility of a technological artefact can vary across user groups.

Each subtype of technological understanding is characteristically associated with a particular context of engagement or 'use', corresponding to the logical distinctions introduced above. These may be described as contexts of design (conceiving or constructing the artefact), practical use (operating it), and innovation (conceptualising new applications for it). This context-sensivity of understanding is also noted by Zhu (2025: p.120). We elaborate on this aspect in a subsequent paper, in which we further explore the three typical contexts we outlined. In the present paper, however, we focus on one of these subtypes—the *via media*—which concerns technological understanding in the design context.

Before proceeding, it is important to emphasise that these accounts, along with the specifications of technological understanding they represent, are *logical options*. In practice, these options often overlap and cannot always be neatly distinguished. For example, the process of thinking up a new relationship between an aim and an artefact can include the process of thinking up the artefact itself. In this scenario, a context of (what we will call) 'innovation' overlaps with a context of 'design'. Similarly, in practice, artefacts can be used for other ends than originally intended, so that a context of 'operation' becomes an instance of 'innovation'. However, for conceptual clarity, it is best to distinguish between different scopes—and thus, subtypes of—technological understanding.

### 3.3. Technological understanding in a context of design

Since 'using a technological artefact' presumes the existence of the artefact itself, it is relevant first to consider a scenario where the artefact has yet to be developed—a situation that mirrors the need to develop a scientific theory in order to achieve scientific understanding.[29] This scenario typically corresponds to a design context.

Kroes (2002, pp. 298-99) characterises the designer's task as turning a description of the intentional aspect of a technological artefact ('functional description') into a description of the physical aspect of the artefact ('structural description'). In our account, the 'functional description' of the artefact(-to-be) corresponds—to some degree[30]—to what we have called the ultimate aim $A$, while the 'structural description' matches our triple $\langle X, P, a\rangle$ (see Section 2.2). The design process thus starts with a description of the ultimate aim,[31] and succeeds if an artefact can be designed that is capable of realising that aim. In other words, turning a functional description into a structural description involves designing a physical structure $X$ that generates a phenomenon $P$ capable of achieving a direct aim $a$, which realises the ultimate aim $A$. This can be reformulated by saying that achieving technological understanding in a design context is a matter of finding an appropriate $t = \langle X, P, a\rangle$ that can realise a specific $A$ (i.e. solving a problem or fulfilling a need).

A key element of this process is the creative step of selecting a phenomenon that can be harnessed to achieve a specific aim. This reveals an asymmetry between scientific and technological understanding. In the process of scientific understanding, the theory-independent question *Why*

[29] This is foreshadowed in De Regt's work because his case studies and his own development of the notion of scientific understanding focus on the understanding that scientists have rather than on understanding by non-scientists (even though his own notion in principle applies to both). (Indeed, his CIT explicitly mentions scientists.) Thus we argue that it is also appropriate for us to begin our account of technological understanding by focussing on the technological understanding that designers have, and in a second step on the understanding by non-designers. We take this second step in De Jong & De Haro (2025b), where we develop our general notion of technological understanding, CTU, in contexts other than the design context.

[30] The intentional aspect that is described in the functional description is, strictly speaking, *part* of the technological artefact in the sense that the direct aim $a$ is inherent to the artefact. However, since the ultimate aim $A$ is external to the technological artefact itself, a description of $A$ does not fully correspond to the functional description of an artefact.

[31] As discussed in Sections 2.2 and 3.2, we consider the ultimate aim to be external to the artefact, and consider formulating it not as a necessary part of designing; rather, design is concerned with shaping an appropriate artefact, and thus formulating the direct aim that adequately achieves the ultimate aim.

$P$? is the starting point. The phenomenon to be explained is not selected in the process of understanding by theory but is given and fixed; scientific understanding is always relative to a predetermined $P$. By contrast, in technological understanding, the phenomenon $P$ serves as a means to an end and can be altered if it fails to meet the technological aim (see also Boon, 2015).

The process of identifying an appropriate phenomenon and designing a physical structure to produce that phenomenon requires cognitive skills. Typically, the 'space of possibilities'—the range of all potential technological artefacts capable of achieving the aim—is not fully known. Thus, design is not merely a selection from predefined options: it requires qualitative strategies and judgement, i.e. *an ability to use knowledge*, to effectively integrate $X$, $P$, and $a$. This capacity for reasoning is what we call 'understanding', as requiring more than 'knowledge'.

In artefactual design, this reasoning involves causal explanation of how the phenomenon is produced by the physical structure. In other words, to be able to produce, control, and manipulate the phenomenon, one must have sufficient knowledge of how it is caused. Hence, the design process inherently builds upon scientific understanding of the relevant phenomena, and more complex or difficult-to-achieve phenomena demand more advanced scientific understanding. In this sense, scientific understanding is—implicitly or explicitly—part of technological understanding. [32]

In the example of quantum computing, the challenge of selecting an appropriate phenomenon $P$ is illustrated by the ongoing exploration of different approaches to quantum computer design, each focussing on distinct quantum mechanical phenomena (NASEM, 2019, pp. 127-30). Despite these efforts, none of the current methods has succeeded in creating a sufficiently well-functioning network of qubits that yields significantly increased computational power. This suggests that scientists and engineers have not yet succeeded in identifying and producing a phenomenon $P$, through a structure $X$, that successfully instantiates $A$. For example, for the superconducting quantum computer it remains a challenge how to measure the state of the qubits quickly, below the decoherence time, but still precisely and with high fidelity. In turn, this may be because the right physical structure $X$, phenomenon $P$, and conditions to achieve the direct aim $a$ of balancing coherence and measurement fidelity in a way that allows scaling, have not yet been found. In other words, the right kind of triple $\langle X, P, a\rangle$ still remains elusive because $P$ and $X$ are not yet intelligible in this context.

Viewing the development of quantum computers through the lens of technological understanding frames this challenge as an epistemic problem of incomplete technological understanding. The central issue in designing quantum computers is not the scientific question of *Why* $P$? but the technological one of *Which* $X$ *for which* $P$ *for* $a$? Answering this question is essential for designing the artefact $t$. This example underscores how selecting a suitable phenomenon in a design context requires more than knowledge—it requires technological understanding to integrate $X$, $P$, and $a$ effectively, such that $A$ is achieved.

To summarise: Technological understanding is demonstrated in recognising how an aim can be realised by using a technological artefact. This ability can be specified as the ability to design an appropriate technological artefact. The cognitive skill involved in selecting a phenomenon $P$ from among possible alternatives, and designing a physical structure $X$ that can generate $P$ to fulfil the direct aim $a$, which in turn embodies the ultimate aim $A$, characterises technological understanding in a design context. Thus successful artefactual design always demonstrates technological understanding, though technological understanding itself is not limited to design (since alternative specifications for 'using a technological artefact' are possible).

[32] This is consistent with the work of Boon, who also highlights the role of understanding physical phenomena such that 'instrument […] can be built, controlled, created, calibrated and/or otherwise manipulated.' (Boon, 2015: p.73; 2017; p. 2)

Accordingly, we can formulate the Criterion for Technological Understanding (CTU) in a design context as follows:

> **CTU(design)**: An aim $A$ is technologically understood if it can be realised by *designing* a technological artefact $t$, i.e. by constructing a physical object $X$ that, by producing a phenomenon $P$ through a physical object $X$, achieves a direct aim a that embodies the ultimate aim $A$

In a design context too, technological understanding hinges on the intelligibility of the artefact. More precisely, the ability to successfully design a technological artefact *implies* that it must be intelligible. As we discussed in the sections on scientific and technological understanding (Sections 2.1 and 3.1), intelligibility is relative to the (typical) subject and depends on contextual factors. In a design context, this means that the artefact must be intelligible to the designer in terms of how its physical structure $X$ and the phenomenon it produces $P$ are capable of achieving the desired aim $a$, embodying the ultimate aim $A$. In other words, the success of the design depends on the designer's ability to establish these relations. Therefore, the necessary level of intelligibility is tied to the designer's capacity to reason about how changes in the artefact's physical structure will lead to corresponding changes in the phenomenon produced, and ultimately, in the fulfilment of the aim. Thus, intelligibility of the artefact in a design context is about understanding the qualitative relationship between $X$ and $P$, and how that relationship can be controlled or manipulated to achieve the intended result.

Table 1 summarises our account of technological understanding up to this point. Rows 1 and 2 capture the analogy between (the function of) a scientific theory and a technological artefact, as discussed in Section 2.3. Building on this analogy, rows 3 and 4 draw comparisons between the types of understanding required for, and demonstrated by, the skilled use of these tools. Row 4a details that the use of a theory or an artefact implies their selection or, where necessary, their development or design. Row 4b highlights that the design of an artefact requires the additional step of selecting a suitable phenomenon, a step that has no equivalent in scientific understanding.

| | Unit of analysis | Scientific theory $\boldsymbol{T}$ | Technological artefact $\boldsymbol{t}$ |
|---|---|---|---|
| *1* | Aim | Explain a phenomenon $P$<br>*Answering 'Why $P$?'* | Realise a practical aim $A$<br>*Solving a problem or meeting a need* |
| *2* | How to achieve the aim | Use a scientific theory $T$<br>*Derive $P$ using $T$* | Use a technological artefact $t$<br>*Generate $P$ using $X$* |
| *3* | Type of understanding | Scientific understanding of $P$ | Technological understanding of $A$ |
| *4* | How to achieve understanding | Explain $P$ using $T$<br>*Construct an interpretation of* $P$ *by using* $T$ | Realise $A$ using $t$<br>*Achieve a direct aim* $a$ *by using* $t$, *that embodies the ultimate aim* $A$ |
| *4a* | Step to understanding | Select or develop $T$ | Select or design $t$ |
| *4b* | Step to understanding | | Select $P$ and find $X$ capable of generating $P$ |

**Table 1** Analogy between scientific theories and technological artefacts, and by extension, between scientific understanding and technological understanding; the blank space indicates the disanalogy.

# 4. Conclusion

The aim of this paper has been to develop an account of understanding in relation to technology. We proposed the concept of *technological understanding*, which is the kind of understanding that is involved in the design and use of technological artefacts. We derived this notion by extending De Regt's (2017) notion of scientific understanding into the domain of technology, and employing an analogy between scientific theories and technological artefacts as our guide. We argued that, just as science seeks to understand and explain natural phenomena through the formulation and use of theories, technology seeks to solve practical problems through the design and use of artefacts.

Drawing on this analogy, we argued that the design and use of a technological artefact involve a type of understanding: it requires the cognitive skill to recognise how an artefact's operation can achieve a particular aim. Just as scientific understanding involves using a theory to explain a phenomenon, technological understanding involves using a technological artefact to realise a practical aim.

Our account of technological understanding parallels scientific understanding, since both involve the cognitive ability to use knowledge effectively—either to achieve explanations (in science) or practical outcomes (in technology). In both cases, such understanding relies on the intelligibility of the respective tools—whether theories or artefacts—to the subject using them.

We further specified the general notion of technological understanding for a design context, arguing that, in this context, technological understanding entails the ability to realise a practical aim by designing an appropriate technological artefact. Specifically, it involves the ability to use knowledge to produce a phenomenon through a physical structure that achieves the aim. Thus, technological understanding is a central epistemic aim in the practice of design.

Our account of technological understanding provides a novel and comprehensive perspective on what understanding entails in relation to technology. It specifically explicates the cognitive skill that is involved in the design and use of technological artefacts, highlighting the epistemic dimension of these activities. To further advance the concept of technological understanding, additional research is required to clarify what 'the ability to use a technological artefact' amounts to in other contexts, including the context of practical use and broader innovation. Such a contextual, multifaceted account of technological understanding not only provides a framework for efforts to improve and assess technological understanding in specific contexts, but also promotes a more diverse view on expertise. We explore these themes in a follow-up paper.